\documentclass[twocolumn,showpacs,preprintnumbers,amsmath,amssymb]{revtex4}
\usepackage{graphicx}% Include figure files
\usepackage{dcolumn}% Align table columns on decimal point
\usepackage{bm}% bold math

\begin{document}

\preprint{APS/123-QED}

\title{Experimental Study of the Role of Atomic Interactions on Quantum Transport}

\author{K. Henderson}
\author{H. Kelkar}
\author{B. Guti\'{e}rrez-Medina}
\altaffiliation[Present address: ]{Stanford University, Stanford, California.}
\author{T. C. Li}
\author{M. G. Raizen}
\affiliation{%
Center for Nonlinear Dynamics and Department of Physics,\\
The University of Texas at Austin, Austin, TX 78712, USA}%

\date{\today}

\begin{abstract}
We report an experimental study of quantum transport for atoms confined in a periodic potential and compare between thermal and BEC initial conditions.  We observe ballistic transport for all values of well depth and initial conditions, and the measured expansion velocity for thermal atoms is in excellent agreement with a single-particle model.  For weak wells, the expansion of the BEC is also in excellent agreement with single-particle theory, using an effective temperature.  We observe a crossover to a new regime for the BEC case as the well depth is increased, indicating the importance of interactions on quantum transport.

\end{abstract}

\pacs{03.75.Kk, 03.75.Lm}% PACS, the Physics and Astronomy
                             % Classification Scheme.
%\keywords{Suggested keywords}%Use showkeys class option if keyword
                              %display desired
\maketitle

In recent years, the topic of quantum transport of weakly interacting particles has attracted increasing attention both theoretically and experimentally \cite{Phillips1, Arimondo1, Inguscio4, Oberthaler1, Stringari1}.  Likewise, experimental efforts to explore the dynamics of BEC atoms in lower dimensions \cite{Ertmer1, Ketterle2, Arimondo2} have also received considerable attention.  Well-tailored magnetic or optical trapping potentials have proven to be very useful tools for reproducing conditions that closely approximate quasi-1D or -2D systems.  In particular, investigations tied to the dynamics of Bose-Einstein condensates (BECs) in quasi-1D optical lattices have proven to be highly non-trivial.  It has already been shown that transport phenomena of weakly interacting particles can play a major role in modifying the dynamics of Bloch oscillations \cite{Arimondo1}, coherence \cite{Inguscio2}, superfluidity \cite{Inguscio1}, nonlinear self-trapping \cite{Oberthaler2}, and inhibited transport \cite{Phillips2}.  

In this Letter, we study the effect of interactions on the quantum transport of bosonic atoms in an optical lattice in regimes not studied previously and find qualitatively new behavior.   We measure the expansion rate as a function of optical lattice depth for noninteracting (thermal) and weakly interacting atoms (BECs).  We understand the expansion of thermal atoms using band theory for a single, non-interacting particle.  In the case of a BEC, we assign an {\it effective temperature} to the BEC and compare that with the theory for thermal atoms at that effective temperature. The agreement is good for low well depths where the expansion rate is larger than the velocity of sound in the BEC. For higher well depths however, single-particle theory proves insufficient to explain the experimental results.

Our experimental sequence begins with a Zeeman-slower loaded magneto-optical trap of $2 \times 10^9$ sodium atoms.  The atoms are optically pumped into the ${\rm F} = 1$, $m_F = -1$ state and transfered to a 'cloverleaf' type Ioffe-Pritchard type magnetic trap \cite{Ketterle1}, with trapping frequencies of $\omega_\perp\;=\;2 \pi \times 324\;$Hz and $\omega_{\rm z}\;=\;2 \pi \times 20\;$Hz in the radial and axial directions respectively.  After 20\;s of RF evaporation we create pure Bose-Einstein condensates (BECs) with $\sim 5 \times 10^6$ atoms.  (By changing the final RF evaporation frequency we can also create  thermal atom samples with no discernible trace of BEC atoms.)

A Nd:YAG (1064\;nm) optical tweezer, with a spot size ($1/e^2$) of $180\;\mu$m and power of $6.35\;$W at the location of the atoms, is then adiabatically ($100\;$ms) ramped on around the BEC atoms.  While being held by the optical tweezer, the original anisotropic magnetic trap is transformed in three stages into a quasi-1D magnetic waveguide \cite{Ertmer1}.  First, the curvature field (axial direction) is turned off in $20\;$ms so that we create a flat waveguide.  Next, a gradient field in the axial direction is adjusted in $50\;$ms to compensate for any tilt along the axial direction \cite{estimate1}.  Finally, atoms are allowed to equilibrate during $500\;$ms before the start of expansion in the waveguide.  This entire procedure has no measurable heating effect and can efficiently transfer more than $2 \times 10^6$ of nearly pure BEC atoms.  The trap frequencies in this hybrid trap are $2 \pi \times 317(1)\;$Hz and $2 \pi \times 75(1)\;$Hz in the radial and axial directions respectively.  Lifetimes in this hybrid trap have been measured to be over $10\;$s.

Great effort was invested to optimize the flatness of the waveguide.  To this extent, both the tilt and flatness of the magnetic waveguide were studied extensively.  Atoms were released from the optical tweezer and both their center of mass motion as well as their spatial extent were measured.  This procedure was repeated for different initial positions along the waveguide, covering a total distance of $750\;\mu$m ($\pm\;375\;\mu$m from the center of the trap).  We compared expansion rates for each initial position and found them to be indistinguishable.  By displacing the cloud in the axial direction, we tracked one quarter of an oscillation from which we estimated the frequency to be $\omega_{\rm z} = 2 \pi \times 820\;$mHz.  We also verified that the density profiles remained fixed in the radial direction throughout an entire expansion \cite{estimate2}.  All measurements were ultimately limited by the resolution of our imaging system ($\sim 5\;\mu$m) and the finite current control of the linear gradient and curvature coils.

The final transition of our experimental sequence is made as atoms are loaded into the optical lattice while being held by the optical tweezer.  We ramp on a periodic potential in $60\;$ms along the axis of the waveguide.  This length of time was chosen to be sufficiently long as to minimize heating.  The periodic potential, which has a form $V(z) = V_0 \sin^2(k_{\rm L}z)$, is created by a repulsive standing wave of far off-resonant light ($\lambda = 532\;$nm) whose waist at the location of the atoms is $120\;\mu$m.  Well depths as high as $s \equiv V_0/E_{\rm R} = 18$ can be attained.  Here $E_{\rm R} = \frac{\hbar^2 k_{\rm L}^{2}}{2 m}$ is the recoil energy and $k_{\rm L} = 2 \pi /\lambda$.  For our experimental parameters, the spontaneous scattering rate can be neglected as atoms expand in the optical lattice.  Also, the Rayleigh length ($8.5\;$cm) of the optical lattice beams is substantially larger than the extent of expansion ($\sim 500\;\mu$m) and therefore does not cause significant variation in the well-depths.

\begin{figure}
\includegraphics[width=8.5cm,height=6cm]{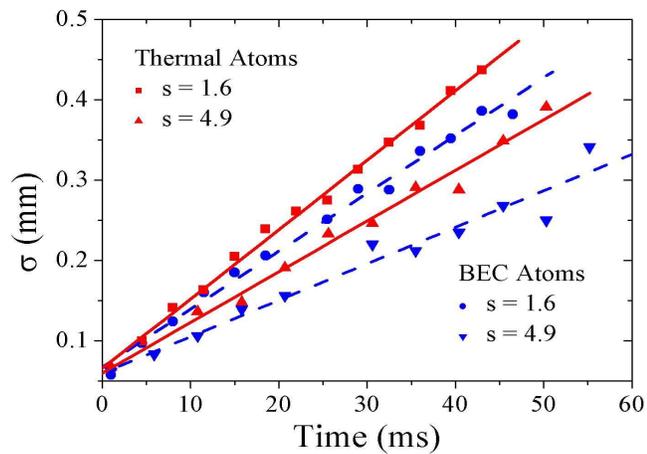}% Here is how to import EPS art
\caption{\label{fig:1}  (color online).  The time evolution of the rms size $\sigma$ is shown for two cases of periodic potential heights, $s = 1.6, 4.9$.  For this plot the thermal atoms are $T=0.16\;T_{\rm R}$.  Dotted lines (BEC) and solid lines (thermal) are fits to the data.  BEC atoms ({\large{$\bullet$}}\;$ 1.6$, {\small{$\blacktriangledown$}}\;$ 4.9$) and thermal atoms ({\tiny{$\blacksquare$}}\;$ 1.6$, {\small{$\blacktriangle$}}\;$ 4.9$).}
\end{figure}

The release from the optical tweezer into the periodic potential is done rapidly ($< 10\;\mu$s).  We verify, from time-of-flight measurements, that this release retains nearly $0.9(1)$ of the condensate fraction.  We also rule out significant heating caused by the optical lattice as determined by holding the atoms in the hybrid trap in the presence of the optical lattice with a maximum well depth of $s = 18$.  Condensate fractions as high as $0.8(1)$ are measured for holding times of $500\;$ms.

After a variable expansion time in the waveguide, all trapping fields are turned off and the atomic distribution is detected by absorption imaging after $3\;$ms of free expansion, from which we measure the axial rms width $\sigma$.  Fig.\;\ref{fig:1} shows a sample of the long time expansions of thermal and BEC atoms for optical potential depths $s = 1.6, 4.9$.  It clearly depicts that thermal atoms with $T=0.16\;T_{\rm R}\;({\rm here},\;T_{\rm R} \equiv 2 E_{\rm R}/k_{\rm B})$ have a faster expansion rate than BEC atoms for a given lattice well depth. 

\begin{figure}
\includegraphics[width=8.5cm,height=6cm]{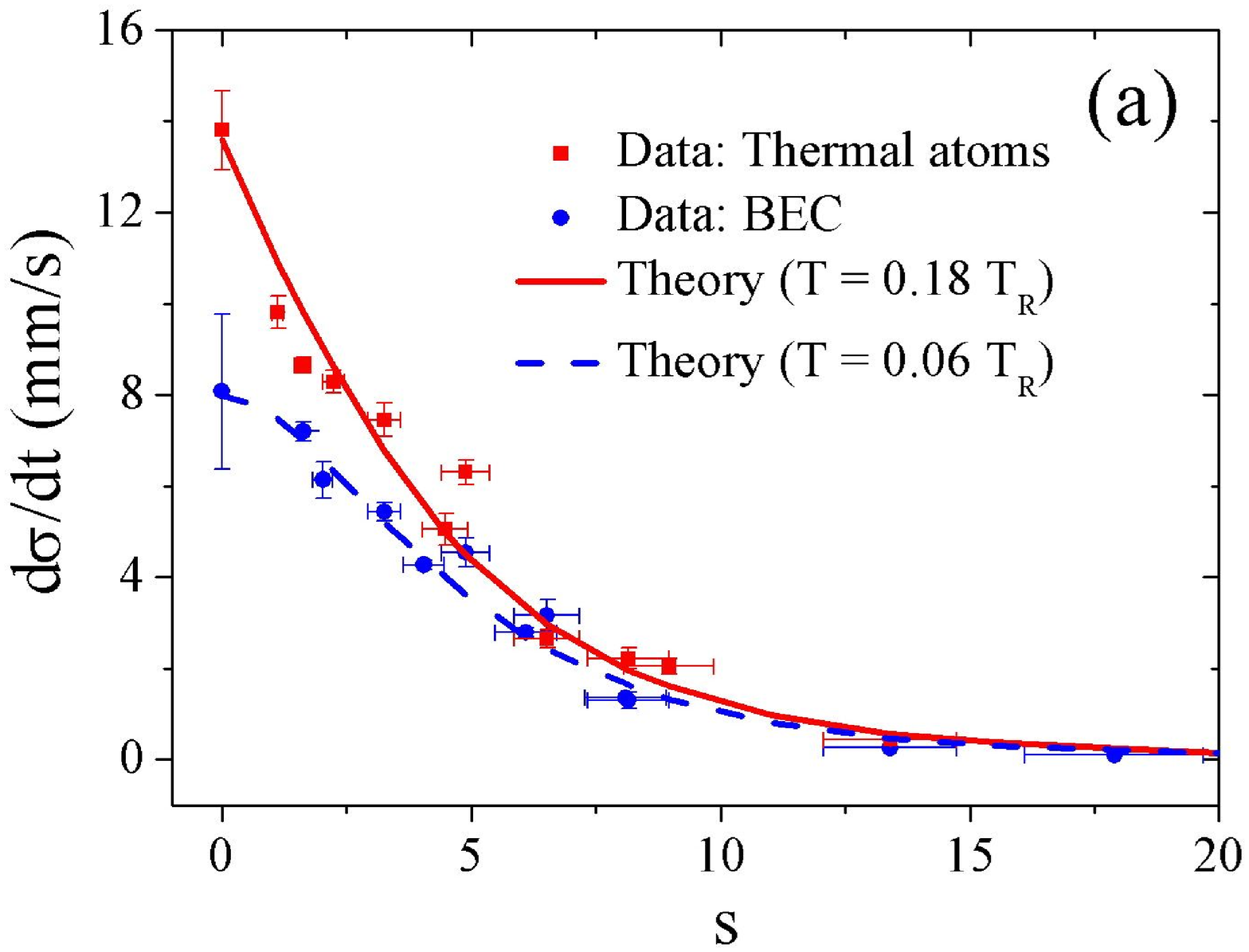}\\ %Here is how to import EPS art
\includegraphics[width=8.5cm,height=6cm]{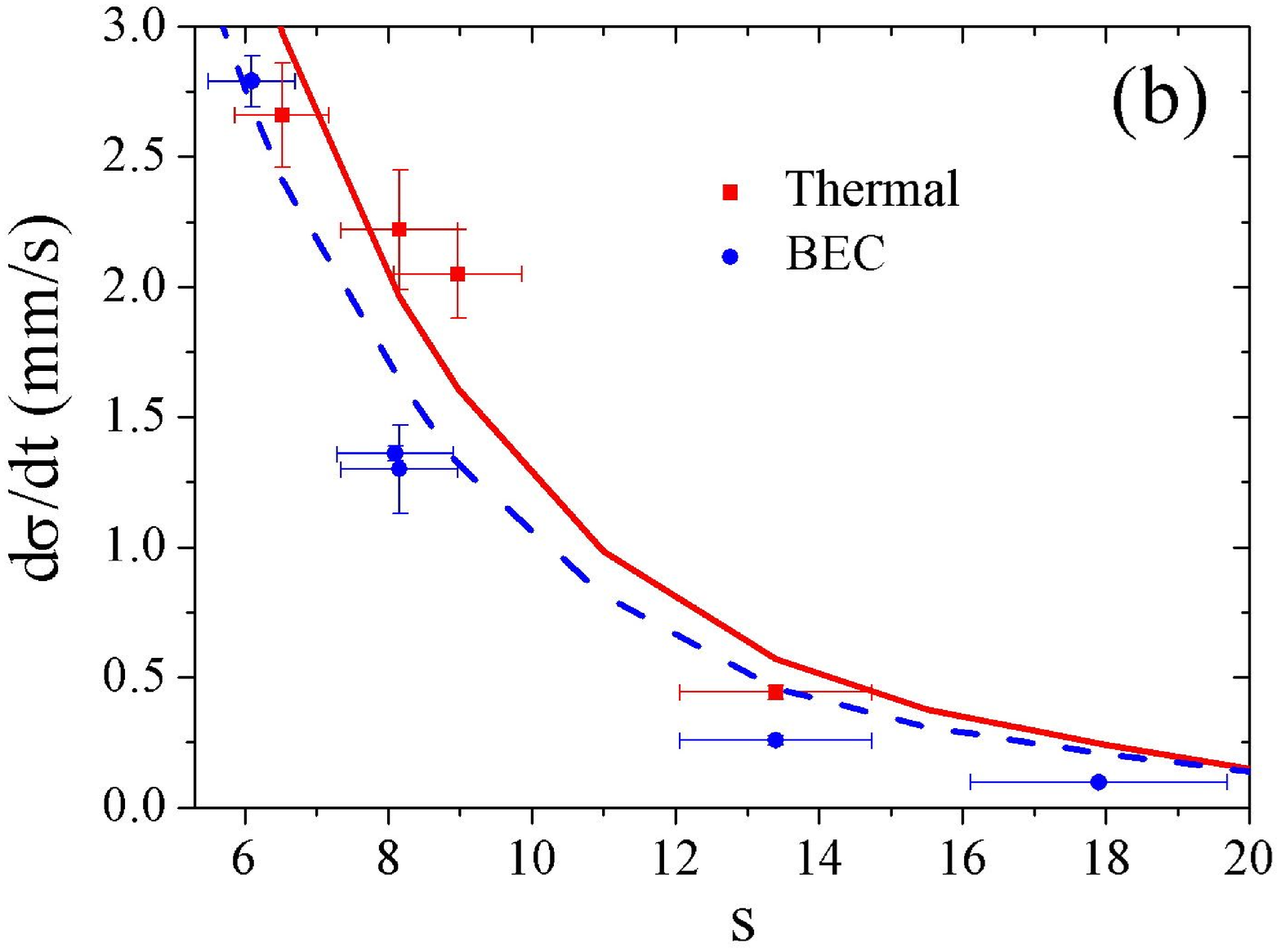} %Here is how to import EPS art
\caption{\label{fig:2}  (color online).  (a) Expansion rates for various periodic potential heights are shown for both thermal atoms ({\tiny{$\blacksquare$}}) and BEC atoms ({\large{$\bullet$}}).  Error bars for $d\sigma/dt$ represent one standard deviation.  Error bars for the optical well depths represent a systematic uncertainty of $\pm\;10\;$\%.  The solid line (thermal) and dashed line (BEC) are theoretical predictions for our experimental parameters based on single-particle band structure theory (see text).  (b) Shows plot (a) in detail for high well depths.}
\end{figure}

Fig.\;\ref{fig:2}(a) shows the dependence of expansion rate ($d\sigma/dt$) on the lattice depth.  The atom number used for the BEC data is $N\;=\;1.7(4) \times 10^6$ (solid squares) and for thermal atom data is $N\;=\;0.9(2) \times 10^6$ (solid circles).  $s = 0$ corresponds to free expansion in the waveguide.  In the absence of the periodic potential, thermal atoms have a linear expansion rate of $13.8(9)\;$mm/s and BEC atoms have a linear expansion rate of $8.1(1.7)\;$mm/s.  The expansion rate for thermal atoms in the waveguide is consistent with their time of flight measured temperature, which is $T=0.16\;T_{\rm R}$.  It is also verified that expansion rates in the waveguide are proportional to the square root of the initial temperature.

For our initial conditions, the condensate atoms in the trap are described accurately by the Thomas-Fermi approximation.  This approximation to the Gross-Pitaevksii equation neglects kinetic energy and predicts the interaction energy per atom to be $E_{\rm int}/N = (2/7)\mu$, where $\mu \propto N^{2/5}$ is the chemical potential of the BEC \cite{Pitaevskii1}.  Prior to release from the optical tweezer our BEC has a chemical potential of $\mu \sim 7\;$kHz for $N \sim 1.7 \times 10^6$ atoms.  For free expansions in the waveguide we measure the kinetic energy in the axial direction and find $E_{\rm kin}/N \cong (1.9/7)\mu$.  Although the dynamics of the first phase of expansion are very complex, what we measure implies that only the interaction energy is released \cite{Ertmer1}.  The potential energy does not convert into the kinetic energy in the axial direction: a result which is the same for thermal atoms.  This analysis also agrees with our observation that the density profiles do not change in the radial direction throughout the entire expansion.

%  THEORY: THERMAL

For finite well depths ($s>0$) single-particle band structure theory is used to describe the ballistic expansion of thermal atoms.  This theory is based on a standard 1-D periodic lattice model and has no adjustable parameters.  We assume the thermal atoms initially satisfy a Maxwell-Boltzmann distribution.  We also assume that their distribution in $k$-space remains unchanged while loading the atoms (adiabatically) into the optical lattice.  The dynamics, then, of an individual atom inside the optical lattice are governed by the semiclassical equations of motion:
\vspace{-2 mm}
\begin{eqnarray*}
\displaystyle \frac{d(\hbar k)}{dt} & = &  0 \\
\displaystyle v_{\rm k} (k) & = & \frac{1}{\hbar}\frac{dE(k)}{dk}
\end{eqnarray*}
where $v_{\rm k}$ is the velocity of an atom in the periodic potential, $k$ is quasi-momentum, and $E(k)$ is the energy per particle in the presence of the periodic potential.

In the absence of any external forces, the evolution of the atoms' distribution in the periodic potential can be written as
\vspace{-2 mm}
\begin{eqnarray*}
\displaystyle f(z,t) & = & N \frac{\hbar \omega_{\rm z}}{2 \pi k_{\rm B} T} \times \\
& & \displaystyle \int e^{-m \omega_{\rm z}^2 (z-v_{\rm k}t)^2/(2 k_{\rm B} T)} e^{-(\hbar k)^2/(2 m k_{\rm B}T)}\ dk
\end{eqnarray*}
where $\omega_{\rm z} = 2 \pi \times 75\;$Hz is the trapping frequency of the optical tweezer and $T$ is the temperature of the atoms, which can be deduced from the free expansion in the waveguide.  By calculating $f(z,t)$ we can predict both density profiles and expansion rates ($d\sigma/dt$).  As depicted in Fig.\;\ref{fig:2}(a), this model is in good agreement with the data collected for expansion rates for thermal atoms.

In the case of interacting atoms, we expect this simple model to provide inaccurate results.  For relatively low well depths ($s \lesssim 6$), surprisingly, we find that the expansion rates versus well depth for a BEC fit very well using the theory for noninteracting atoms.  For high well depths, however, we observe that the theory clearly deviates from the experimental results.

\begin{figure}
\includegraphics[width=8.25cm,height=4cm]{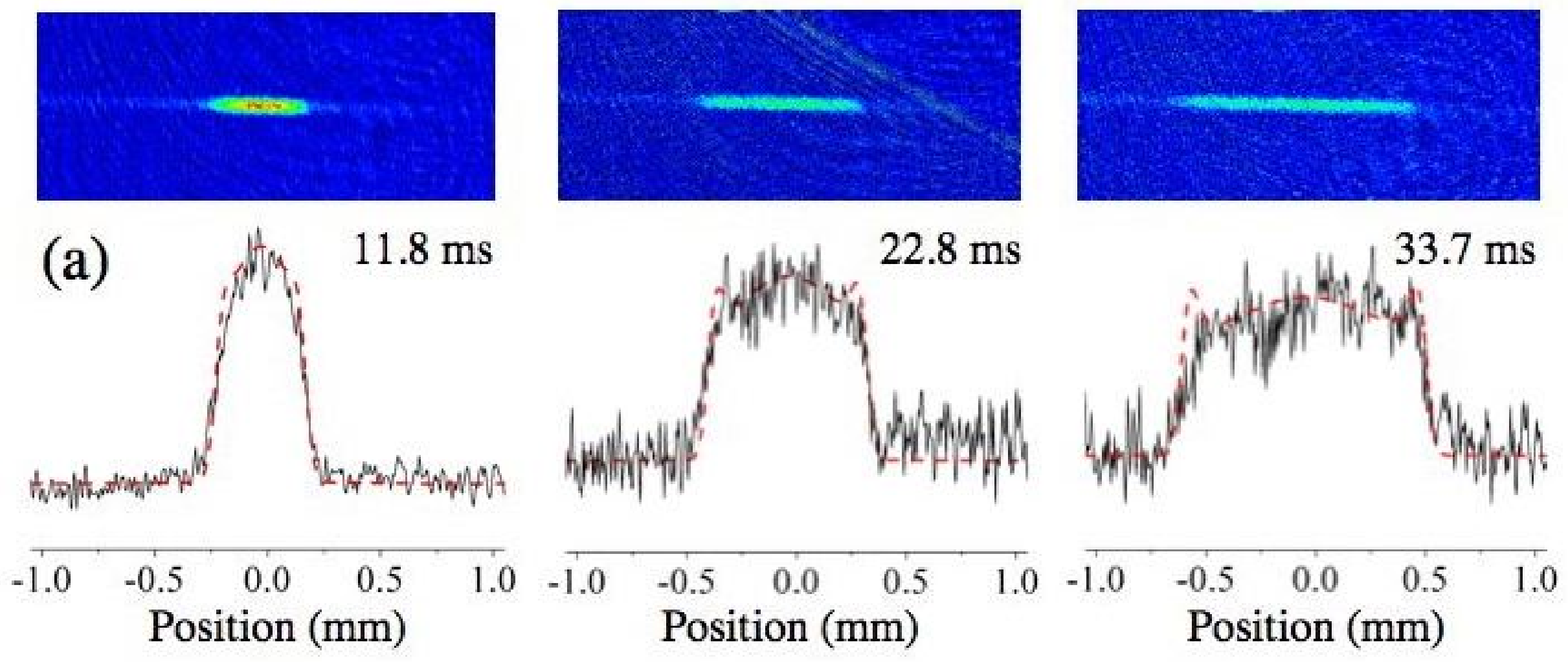}\\% Here is how to import EPS art
\includegraphics[width=8.25cm,height=4cm]{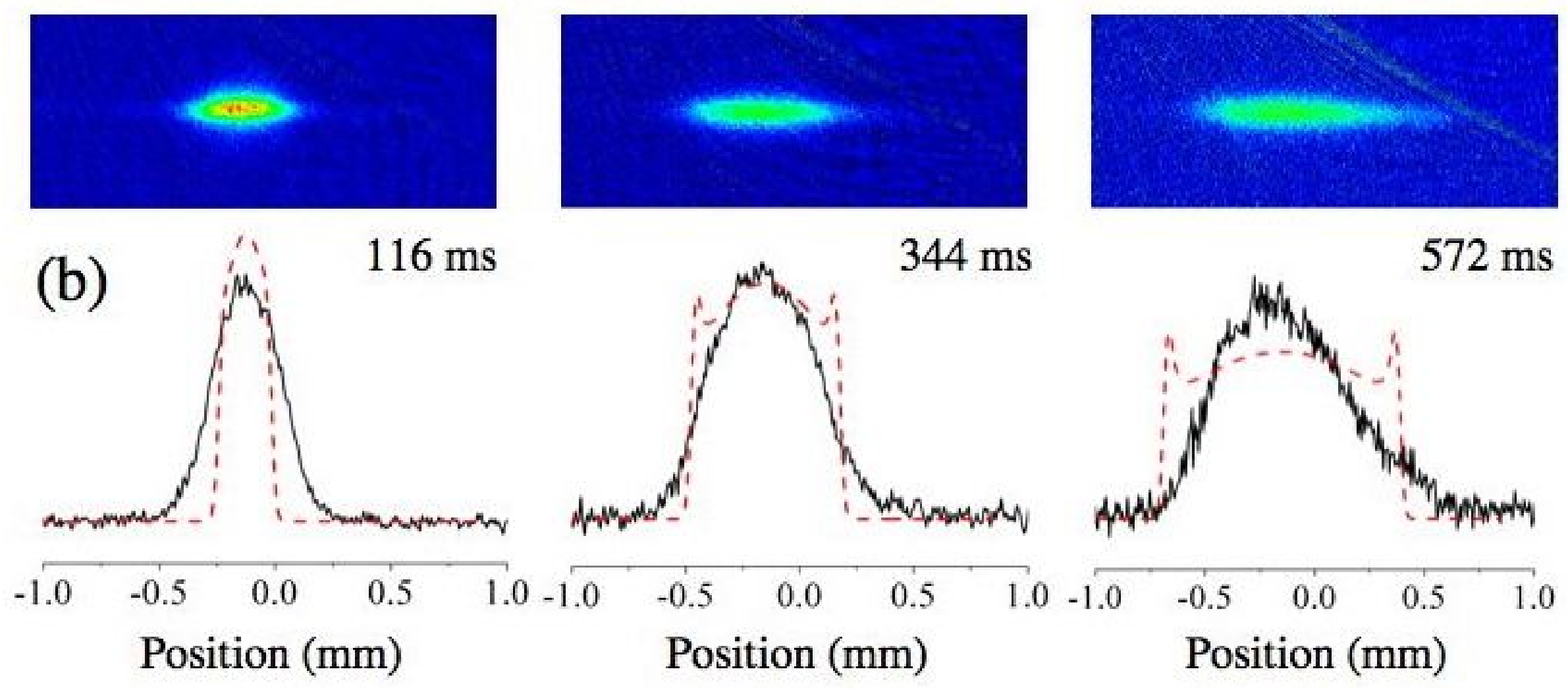}% Here is how to import EPS art
\caption{\label{fig:3}   (color online).  (a) The evolution of density profiles is shown for thermal atoms ($N = 0.54(5) \times 10^6$ atoms, T\;$= 0.18\;T_{\rm R}$) expanding in an optical lattice with $s = 2.25$. (b) The evolution of density profiles is shown for BEC atoms ($N = 1.8(3) \times 10^6$ atoms, with an effective temperature T\;$=0.06\;T_{\rm R}$) expanding in an optical lattice with $s = 13.4$.  For both plots the dotted lines are theoretical density profiles based on single-particle band structure calculations (see text).}
\end{figure}

For low well depths, the rate of expansion exceeds the estimated velocity of sound ($\sim 3\;$mm/s) for our typical initial peak densities of BEC atoms ($\sim 8 \times 10^{13}\;/{\rm cm}^3$) \cite{estimate3}.  During such fast expansions, the BEC as a whole, no longer behaves as a single entity, but as a collection of individual atoms.  Interactions, we observe, seem to play very little role for such expansions.  As suggested in Ref.\;\cite{Shlyapnikov1}, we can thus ascribe an {\it effective} temperature to the expanding atoms which is obtained by fitting the data point $s = 0$ for the BEC atom data (in this case $T = 0.06 T_{\rm R}$).

For higher well depths the interaction-driven dynamics of the BEC atoms is more complex than what single-particle theory can predict.  From Fig.\;\ref{fig:2}(b) we observe that there is a trend for BEC atoms to expand much less than the rate predicted by single-particle band structure model.  In the case of the expansion rate measured for $s = 17.9$, the rate was found to be nearly half of the expected value with errors less $\pm\;10\;$\% in the uncertainty of expansion.

As shown in Fig.\;\ref{fig:3}, we also compare the density profiles of thermal atoms to those of a BEC.  We  use the same single-particle band structure theory to predict a profile and compare it with experimental data.  According to the theoretical model, the development of sharp edges in the density profiles is a consequence of the maximum allowed velocity in the lowest band of the optical lattice.  For sufficiently cold atoms (\;$\lesssim0.1\;E_{\rm R}/k_{\rm B}$) density profile edges do not emerge for any well depth.  In contrast, for hotter atoms, density profiles like those seen in Fig.\;\ref{fig:3}(a) are typical.  At high well depths the maximum allowed velocity becomes smaller than for low well depths, therefore atoms must be allowed to expand for longer times in order for the appearance of density profile edges to be observable.

The density profiles for thermal atoms are in excellent agreement with theory, as shown in Fig.\;\ref{fig:3}(a).  In the case of BEC atoms, the density profiles are not in good agreement with the theory predicted profiles for high well depths.  As shown in Fig.\;\ref{fig:3}(b) we have observed the long time evolution of the density profile for BEC atoms.  In $100\;$ms the atoms have expanded from an initial rms waist of $90\;\mu$m to $120\;\mu$m.  The waist of the profile is noticeably larger than the predicted value using the theoretical model.  This discrepancy is due, in part, to the initial interaction strength of the BEC atoms.  It is well known that for large atom number a Thomas-Fermi distribution will acquire a profile that differs in size from that of thermal atoms \cite{Pitaevskii1,Wieman1}.  For longer times, however, we have observed that the atoms expand less than the model predicts.  We {\it do not} see transport stop for any well depth, i.e., the sharp edges continue to grow with the maximum velocity allowed in the lowest band.  Likewise, we do not believe atoms stop or slow down due to axial trapping frequency because we have noted that their expansion still grows linearly for times greater than $800\;$ms.
 
The fact that expansion rates for BEC atoms for low well depths can be successfully modeled using a single-particle theory demonstrates the extent to which BEC interactions have ceased to participate in the dynamical role of transport through the optical lattice.  This result may have a simple explanation.  The density and collision rate for BEC atoms drops by nearly a factor of three when atoms are allowed to expand for just $10\;$ms in a low well depth ($s \sim 2$ -- $6$) optical lattice.  As non-linear phenomena associated BEC atoms is inextricably linked to their interactions, a sharp drop in density will likely mitigate mean-field effects \cite{Shlyapnikov1}.

In summary, we observe a crossover from interacting to noninteracting dynamics in a variety of optical lattices.  This work should stimulate further theoretical work on the transport properties of BEC atoms and thermal induced decoherence \cite{Stringari3}.

The authors would like to acknowledge support from the Sid W.
Richardson Foundation, the National Science Foundation, the R. A.
Welch Foundation.

\bibliography{Bloch}% Produces the bibliography via BibTeX.

\begin{thebibliography}{21}
\bibitem[{est({\natexlab{a}})}]{estimate1}
\bibinfo{note}{By observing the center of mass motion, we found that the
  acceleration along the axial direction of the waveguide was less than
  $250\;\mu m/s^2$ over $750\;\mu$m of the waveguide.}
\bibitem[{est({\natexlab{b}})}]{estimate2}
\bibinfo{note}{For different optical lattice well-depths, $s$, the spatial
  extent in the radial direction does change accordingly. However, for each of
  these cases, respectively, the extent remains fixed.}
\bibitem[{est({\natexlab{c}})}]{estimate3}
\bibinfo{note}{For ground state BEC atoms, the velocity of sound is $c_{\rm s}
  = \sqrt{n g/m}$ where $n$ is the density and $g = (4 \pi \hbar^2 a_s)/m$. In
  the presence of a 1D optical lattice $c_{\rm s}$ decreases as a function of
  higher well depths, $s$. See Ref. \cite{Stringari2}. Expansions into the
  optical lattice ultimately cause a decrease in density and therefore a
  further decrease in sound velocity.}
\end{thebibliography}

\end{document}